\documentclass[preprint]{article}
\usepackage{authblk}
\usepackage{graphicx}% Include Figure files
\usepackage{dcolumn}% Align table columns on decimal point
\usepackage{bm}% bold math
\usepackage{amsmath}
\usepackage{amssymb}
\usepackage{multirow}
\usepackage{caption}
\usepackage{subcaption}
\usepackage{epstopdf}
\usepackage{hyperref}
\usepackage{cite}
\begin{document}
\title{\bf Constraining $f(Q,\mathcal{L}_{m})$ gravity with redshift-dependent pressure: Insights from observational probes}
\author[]{Amit Samaddar\thanks{samaddaramit4@gmail.com}}
\author[]{S. Surendra Singh\thanks{ssuren.mu@gmail.com}}
\affil[]{Department of Mathematics, National Institute of Technology Manipur, Imphal-795004,India.}

\maketitle

\begin{abstract}
 We explore the late time cosmological dynamics of the Universe within the framework of $f(Q,\mathcal{L}_{m})$ gravity by considering the specific form $f(Q, \mathcal{L}_m)=-Q+2\mathcal{L}_m+\gamma$. To describe the cosmic pressure evolution, a redshift dependent parametrization of $p(z)=\alpha+\frac{\beta z}{1+z}$ is introduced. MCMC analysis is performed using a combined datasets from Hubble ($46$ points), BAO ($15$ points including DESI DR2) and Pantheon$+$ ($1701$ SNe Ia), the model parameters are constrained as $H_{0}=67.9476^{+0.7534}_{-0.7523}$ (km/s/Mpc), $\alpha=-0.0002^{+0.0211}_{-0.0208}$, $\beta=-0.0001^{+0.0410}_{-0.0404}$ and $\gamma=0.0002^{+0.0599}_{-0.0602}$. The model predicts a transition from deceleration to acceleration at $z_{tr} \approx 0.493$ with present values $q_{0}=-0.255$ and $\omega_{0}=-0.9001$. The evolution of energy density and pressure aligns with observational expectations. An analysis of energy conditions shows that NEC and DEC are satisfied, while SEC is violated, consistent with late time acceleration. Moreover, the slow roll parameters $\epsilon_{1}$ and $\epsilon_{2}$ confirm a smooth inflationary regime. These results demonstrate the capability of the model to unify early Universe inflation with the current phase of cosmic acceleration.
\end{abstract}


\section{Introduction}\label{sec1}
\hspace{0.5cm} In the last twenty years, extensive and highly precise cosmological data—from Type Ia supernovae observations \cite{Riess98,Per99}, measurements of the Cosmic Microwave Background (CMB) \cite{Planck2018} and large-scale structure mappings \cite{Alam2021}—have consistently indicated that the Universe is expanding at an accelerating rate. This unexpected behavior, which remains unexplained within the framework of General Relativity by considering only ordinary matter and energy, is commonly attributed to dark energy. This enigmatic form of energy is believed to account for nearly $70\%$ of the Universe’s total energy density. Although the $\Lambda$CDM model offers the most straightforward and widely accepted explanation for the Universe’s accelerated expansion, it is not without significant theoretical difficulties. One major concern is the cosmological constant problem, arising from the vast mismatch between the observed value of $\Lambda$ and theoretical predictions from quantum field theory. Another challenge is the cosmic coincidence problem, which raises the question of why the energy densities of dark energy and matter are comparable in the current epoch \cite{Steven89,Carroll2001}. Moreover, the persistent Hubble tension—referring to the significant difference between the locally measured and early-Universe inferred values of the Hubble constant $H_0$—has intensified doubts about the adequacy of the standard $\Lambda$CDM framework \cite{Verde2019}.

In response to these unresolved issues, researchers have explored various extensions of General Relativity known as modified gravity theories, which aim to explain the observed cosmic acceleration without introducing unknown dark energy components. Among these alternatives, $f(R)$ gravity has gained attention, where the gravitational action is generalized to a function of the Ricci scalar $R$, providing a purely geometric source for acceleration through modifications to the curvature sector \cite{Soti10,Nojiri11}. Similarly, $f(T)$ gravity represents an approach within teleparallel geometry, which relies on torsion instead of curvature, with the action constructed from the torsion scalar $T$ \cite{Amit24}. Theories such as $f(G)$ gravity extend the action using the Gauss-Bonnet invariant $G$, contributing effectively to both early- and late-time cosmological dynamics \cite{Nojiri05,ADEF10}. Other models like $f(R,L_m)$ and $f(R, T)$ theories incorporate explicit couplings between matter and curvature, enabling direct interactions between the material content of the Universe and the geometry of spacetime \cite{Har11, Bertolami2007}. More recently, attention has shifted toward $f(Q)$ gravity, formulated within symmetric teleparallel geometry, where the non-metricity scalar $Q$ governs gravitational effects in the absence of both curvature and torsion \cite{JB18,JB20}. Each of these approaches offers distinct geometric mechanisms for explaining the Universe’s accelerated expansion.

An intriguing advancement in modified gravity is the $f(Q,\mathcal{L}_{m})$ framework, where the gravitational action depends on both the non-metricity scalar $Q$ and the matter Lagrangian $\mathcal{L}_m$ \cite{K2024, Y2024, Y2025, A2025}. This formulation introduces a non-minimal coupling between matter fields and the geometric properties of spacetime, which leads to an effective interaction that bridges matter and dark energy sectors. Such couplings naturally alter the cosmic expansion history and can impact the growth of large-scale structures, which offers potential explanations for the Universe's late-time acceleration as well as existing cosmological discrepancies. Notably, the resulting field equations are of second-order, which preserves compatibility with General Relativity’s predictive successes, while opening new avenues for exploring the interaction between matter content and the underlying geometry of spacetime.

In cosmological studies, parametrization involves representing critical cosmic quantities—like pressure, energy density, or the equation of state (EoS) parameter—as explicit functions of redshift $z$. This method offers a versatile and often theory-independent framework for describing the Universe's expansion, especially when the true nature of dark energy or modified gravity remains uncertain. A prominent example is the Chevallier–Polarski–Linder (CPL) parametrization, where the EoS parameter is modeled as $\omega(z)=\omega_0+\omega_a\frac{z}{1+z}$ \cite{Chevallier2001, Linder2003}, allowing exploration of evolving dark energy behavior. Such parameterized forms simplify complex cosmological models, condensing late-time acceleration features into a few adjustable parameters that can be effectively constrained using observational data. Consequently, parametrization serves as a powerful diagnostic tool to evaluate alternative cosmological theories and to investigate potential deviations from the conventional $\Lambda$CDM framework \cite{Sahni2008,Singh23,AS25,Alam2024}. To investigate the Universe's accelerated expansion under the framework of $f(Q,\mathcal{L}_{m})$ gravity, we employ a pressure parametrization expressed as a function of redshift. This method enables the modeling of cosmic pressure evolution in a model-independent manner, without relying on any predefined dark energy scenario. Such a strategy offers flexibility in tracing potential departures from standard cosmological behavior. The selected parametrization effectively describes key features of late-time cosmic dynamics, ensuring consistency with diverse observational datasets.

The structure of this paper is organized as follows: In section \ref{sec2}, we briefly review the field equations of $f(Q,\mathcal{L}_m)$ gravity. In section \ref{sec3}, we propose a specific form of the function $f(Q,\mathcal{L}_m)$ along with a redshift-dependent pressure parametrization. In section \ref{sec4}, we outline the statistical framework based on the chi-square minimization technique and employ recent observational datasets including Hubble, BAO, DESI DR2 and Pantheon+ to constrain the model parameters. In section \ref{sec5}, we examine the cosmological dynamics through key kinematic quantities such as the deceleration parameter, energy density, pressure and the equation of state. In section \ref{sec6}, we analyze the energy conditions to assess the physical viability of the model. Section \ref{sec7} investigates the inflationary behavior via slow-roll parameters. Finally, we summarize our findings and discuss their cosmological implications in section \ref{sec8}.
\section{Analytical formulation of the $f(Q,\mathcal{L}_{m})$ gravity model}\label{sec2}
The $f(Q,\mathcal{L}_m)$ theory is a broad generalization of symmetric teleparallel gravity, where gravity is attributed to the nonmetricity scalar $Q$, rather than spacetime curvature or torsion. In this framework, the action is constructed from a function that directly couples the nonmetricity scalar $Q$ with the matter Lagrangian density $\mathcal{L}_m$, and is given by:
\begin{equation}\label{1}
S=\int f(Q,\mathcal{L}_{m})\sqrt{-g}d^{4}x,
\end{equation}
Here, $g$ denotes the determinant of the metric tensor $g_{\mu\nu}$. This action maintains general covariance and represents a wider class of gravitational models by explicitly incorporating the interaction between the geometry of spacetime and the matter content.

Within this framework, the nonmetricity scalar $Q$ acts as the fundamental geometric quantity, constructed from specific contractions of the disformation tensor $L^{\lambda}{}_{\mu\nu}$, which characterizes the extent to which the connection deviates from being metric-compatible. The scalar $Q$ is explicitly defined as:
\begin{equation}\label{2}
Q=-g^{\mu\nu} \left( L^{\alpha}{}_{\beta\mu} L^{\beta}{}_{\nu\alpha} - L^{\alpha}{}_{\beta\alpha} L^{\beta}{}_{\mu\nu} \right),
\end{equation}
The disformation tensor is defined in terms of the nonmetricity tensor as:
\begin{equation}\label{3}
L^\lambda{}_{\mu\nu} = \frac{1}{2} g^{\lambda\sigma} \left( Q_{\nu\mu\sigma} + Q_{\mu\sigma\nu} - Q_{\sigma\mu\nu} \right),
\end{equation}
The non-metricity tensor $Q_{\lambda\mu\nu} \equiv -\nabla_{\lambda} g_{\mu\nu}$ quantifies the deviation of the metric tensor from being covariantly constant with respect to the affine connection $\check{L}^{\lambda}{}_{\mu\nu}$, which is taken to be symmetric and devoid of curvature. Two natural trace vectors emerge from this tensor: 
\begin{equation}\label{4}
Q_{\mu} = Q_{\mu}{}^{\alpha}{}_{\alpha}, \quad 
\tilde{Q}_{\mu} = Q^{\alpha}_{\phantom{\alpha}\mu\alpha},
\end{equation}
are naturally introduced and play crucial roles in forming scalar combinations and deriving the equations of motion. From these, one constructs the superpotential tensor $P^{\lambda}{}_{\mu\nu}$, a fundamental object that mediates the dynamics of the theory: 
\begin{equation}\label{5}
P^{\lambda}{}_{\mu\nu} = \frac{1}{4} \left[ -Q^{\lambda}{}_{\mu\nu} + 2 Q_{(\mu\nu)}^{\;\;\;\;\;\lambda} + Q^{\lambda} g_{\mu\nu} - \tilde{Q}^{\lambda} g_{\mu\nu} - \delta^{\lambda}_{\;(\mu} Q_{\nu)} \right]
\end{equation}
This tensor emerges naturally from the variation of the action and enables the scalar quantity $Q$ to be compactly reformulated as
\begin{equation}\label{6}
Q = -Q_{\lambda\mu\nu} P^{\lambda\mu\nu},
\end{equation}
revealing the geometric relationship between the non-metricity tensor and its associated conjugate quantity.

Taking the variation of the action with respect to the metric leads to the generalized or modified gravitational field equations.
\begin{equation}\label{7}
\frac{2}{\sqrt{-g}} \nabla_{\lambda} \left( \sqrt{-g} f_Q P^{\lambda}{}_{\mu\nu} \right) + f_Q \left( P_{\mu\lambda\sigma} Q_{\nu}{}^{\lambda\sigma} - 2 Q^{\lambda\sigma}{}_{\mu} P_{\lambda\sigma\nu} \right) + \frac{1}{2} f g_{\mu\nu} = \frac{1}{2} f_{\mathcal{L}_m} \left( g_{\mu\nu}\mathcal{L}_m - T_{\mu\nu} \right)
\end{equation}
The partial derivatives of the function $f$ with respect to $Q$ and the matter Lagrangian $\mathcal{L}_m$ are denoted as $f_Q\equiv\frac{\partial f}{\partial Q}$ and $f_{\mathcal{L}_m} \equiv \frac{\partial f}{\partial \mathcal{L}_m}$, respectively. The influence of matter is incorporated via the energy–momentum tensor.
\begin{equation}\label{8}
T_{\mu\nu} = -\frac{2}{\sqrt{-g}} \frac{\delta (\sqrt{-g}\mathcal{L}_m)}{\delta g^{\mu\nu}} = g_{\mu\nu}\mathcal{L}_m - 2 \frac{\partial\mathcal{L}_m}{\partial g^{\mu\nu}}
\end{equation}
Moreover, varying the action with respect to the affine connection gives rise to a separate equation that involves the hypermomentum density.
\begin{equation}\label{9}
\nabla_{\mu} \nabla_{\nu} \left[ 4 \sqrt{-g} f_Q P^{\mu\nu}{}_{\lambda} + H_{\lambda}{}^{\mu\nu} \right] = 0,
\end{equation}
where the hypermomentum tensor is defined as 
\begin{equation}\label{10}
H_{\lambda}{}^{\mu\nu} = \sqrt{-g} f_{\mathcal{L}_m} \frac{\delta\mathcal{L}_m}{\delta Y^{\lambda}{}_{\mu\nu}},
\end{equation}
with $Y^{\lambda}{}_{\mu\nu}$ denoting the distortion tensor that captures how the matter Lagrangian $\mathcal{L}_m$ depends on the affine connection. This reflects the distinguishing feature of $f(Q,\mathcal{L}_{m})$ gravity as a metric-affine theory, where the matter sector is allowed to couple both to the metric and the independent connection.

A remarkable and fundamental aspect of the $f(Q,\mathcal{L}_{m})$ gravity theory is that it allows for the explicit non-conservation of the energy–momentum tensor. In contrast to general relativity—where the covariant divergence of $T_{\mu\nu}$ is zero due to diffeomorphism invariance and the assumption of minimal coupling—this framework permits a direct coupling between matter and the non-metricity scalar via the function $f(Q,\mathcal{L}_{m})$. As a result, energy and momentum are no longer conserved independently within the matter sector, but instead can be transferred to or from the geometric structure of spacetime. This departure from energy–momentum conservation is formally expressed by evaluating the covariant divergence of the modified field equations, resulting in a generalized relation that encapsulates the dynamic exchange between matter and geometry.
\begin{equation}\label{11}
D_{\mu}T^{\mu}{}_{\nu}=\frac{1}{f_{\mathcal{L}_{m}}}\bigg(\frac{2}{\sqrt{-g}}\nabla_{\lambda}\nabla_{\mu}H^{\;\;\lambda\mu}_{\nu}+\nabla_{\mu}A^{\mu}_{\nu}-\nabla_{\mu}\bigg[\frac{1}{\sqrt{-g}}\nabla_{\lambda}
H^{\;\;\lambda\mu}_{\nu}\bigg]\bigg)=B_{\nu}\neq 0.
\end{equation}
The quantity $B_{\nu}$ characterizes the transfer rate of energy and momentum between the matter fields and the geometric background. While it identically vanishes in theories with minimal coupling, it generally acquires nonzero values in $f(Q,\mathcal{L}_{m})$ gravity. Its behavior is highly sensitive to the specific forms of both the function $f(Q,\mathcal{L}_{m})$ and the matter Lagrangian $\mathcal{L}_{m}$, reflecting the depth of their interaction.

To investigate the cosmological consequences of this extended gravity framework, we consider a spatially flat Friedmann–Lema$\hat{i}$tre–Robertson–Walker (FLRW) spacetime, which provides a suitable geometric backdrop for modeling a homogeneous and isotropic Universe.
\begin{equation}\label{12}
ds^2 = -dt^2 + a^2(t) \left( dx^2 + dy^2 + dz^2 \right),
\end{equation}
Here, $a(t)$ represents the scale factor that governs the Universe’s expansion. Within this uniform and direction-independent (homogeneous and isotropic) setting, the expression for the non-metricity scalar becomes considerably simplified, which reflects the underlying symmetries of the FLRW geometry as:
\begin{equation}\label{13}
Q=6H^{2},
\end{equation}
In this context, the Hubble parameter $H=\frac{\dot{a}}{a}$ defines the rate at which the Universe is expanding over time.

To describe the matter sector, we assume it behaves as a perfect fluid, whose energy–momentum tensor takes the standard form, capturing the energy density and pressure contributions in a cosmologically symmetric setup.
\begin{equation}\label{14}
T_{\mu\nu} = (\rho + p) u_\mu u_\nu + p g_{\mu\nu},
\end{equation}
Here, $\rho$ and $p$ denote the energy density and pressure of the fluid, respectively, while $u^\mu$ represents its four-velocity, constrained by the normalization condition $u^\mu u_\mu = -1$. This ensures that the fluid's motion remains timelike, consistent with the principles of relativistic hydrodynamics.

By adopting the FLRW metric outlined in equation (\ref{12}) and integrating the perfect fluid energy–momentum tensor from equation (\ref{14}), applying the variational principle to the extended gravitational action defined in $f(Q,\mathcal{L}_{m})$ gravity yields a modified form of the Friedmann equations. These newly derived equations describe how the Universe expands within this specific theoretical model, as detailed in \cite{Hazarika24}.
\begin{equation}\label{15}
3H^{2}=\frac{1}{4f_{Q}}\bigg[f-f_{\mathcal{L}_{m}}(\rho+\mathcal{L}_{m})\bigg],
 \end{equation}
 \begin{equation}\label{16}
\dot{H}+3H^{2}+\frac{\dot{f_{Q}}}{f_{Q}}H=\frac{1}{4f_{Q}}\bigg[f+f_{\mathcal{L}_{m}}(p-\mathcal{L}_{m})\bigg].
 \end{equation}
 \section{A minimal coupling inspired model}\label{sec3}
 \hspace{0.5cm} To investigate the cosmological effects of $f(Q,\mathcal{L}_{m})$ gravity in a straightforward yet insightful manner, we consider a linear model of the theory, expressed as \cite{Hazarika24}:
\begin{equation}\label{17}
f(Q, \mathcal{L}_m)=-Q+2\mathcal{L}_m+\gamma,
\end{equation}
In this model, $\gamma$ is a constant that acts similarly to an effective cosmological constant, while $\mathcal{L}_m$ represents the matter Lagrangian. This setup provides a foundational case where the interaction between the non-metricity scalar $Q$ and matter is simple yet retains meaningful dynamical effects, allowing for minimal but non-negligible coupling. The choice of this specific functional form is guided by both theoretical soundness and physical applicability. The inclusion of the term $-Q$ directly mirrors the teleparallel equivalent of the Ricci scalar $R$ in symmetric teleparallel gravity. When the function takes the form $f(Q,\mathcal{L}_m)=-Q+2\mathcal{L}_m$, the resulting theory becomes dynamically equivalent to general relativity with minimally coupled matter, ensuring that classical gravitational dynamics are recovered in suitable limits. The constant parameter $\gamma$ plays a role analogous to the cosmological constant $\Lambda$ in GR, but here it emerges naturally from the non-metricity framework, which offers a geometric explanation for cosmic acceleration without resorting to arbitrary constants. Furthermore, this model maintains general covariance and remains analytically manageable, which provides clear insights into cosmological behavior while avoiding unnecessary complications from more elaborate couplings or higher-order corrections.

This family of models—linear in both the non-metricity scalar $Q$ and the matter Lagrangian $\mathcal{L}_m$—has gained attention in recent gravitational research. Notably, \cite{Y2024} explored a specific extension given by $f(Q, \mathcal{L}_m) = -\alpha Q + 2\mathcal{L}_m + \beta$, where $\alpha$ and $\beta$ are constant parameters. In this formulation, $\alpha$ effectively modifies the gravitational coupling strength, while $\beta$ introduces a uniform shift in the energy density. Myrzakulov's analysis showed that such models can successfully account for the observed late-time acceleration of the Universe, doing so without invoking additional scalar fields or altering the standard matter sector—highlighting the potential of this framework to explain cosmic acceleration purely through geometric means.

In this study, we choose the matter Lagrangian as $\mathcal{L}_m = \rho$, which is a natural fit within the cosmological framework where a perfect fluid governs the dynamics. This selection reflects the central role of energy density $\rho$ in driving gravitational behavior. Importantly, opting for $\rho$ instead of $-\rho$ preserves the positivity of the matter sector and ensures that the coupling with geometry remains rooted in energy content. Additionally, this choice results in physically consistent and well-behaved sign conventions within the modified Friedmann equations, making it particularly effective for studying both cosmic acceleration and the interplay between matter and geometry \cite{Ober08,Kavya22,Zhao12}.

To characterize the Universe's late-time dynamics, we adopt a redshift-dependent pressure parametrization given by
\begin{equation}\label{18}
p(z)=\alpha+\frac{\beta z}{1 + z},
\end{equation}
where $\alpha$ and $\beta$ are free parameters of the model, and $z$ denotes the redshift. This functional form, originally proposed by \cite{Zhang2015}, serves as a phenomenological tool to capture variations in pressure across cosmic time, particularly deviations from a constant pressure at low redshifts. It transitions smoothly from $p(z) \to \alpha + \beta$ at early times ($z \to \infty$) to $p(z) \to \alpha$ at the present epoch ($z \to 0$). This behavior makes it a flexible and effective choice for modeling time-evolving dark energy or representing fluid-like behavior in alternative theories of gravity.

To track the evolution of the cosmic fluid, we make use of the energy conservation law, which arises from the covariant conservation of the energy–momentum tensor within the context of a spatially flat FLRW Universe.
\begin{equation}\label{19}
\dot{\rho}+3H(\rho + p)=0,
\end{equation}
By applying the continuity equation in combination with the pressure parametrization defined in equation (\ref{18}), we solve the conservation law to obtain an explicit expression for the $\rho(z)$ as:
\begin{equation}\label{20}
\rho(z)=-\frac{3}{2}(\alpha+\beta)(1+z)+\beta+c(1+z)^{3},
\end{equation}
By employing the model defined in equation (\ref{17}) and incorporating the energy density derived from the continuity relation, we insert these results into the modified Friedmann equation given in equation (\ref{15}).
\begin{equation}\label{21}
H^{2}=\frac{1}{3}\bigg[-\frac{3}{2}(\alpha+\beta)(1+z)+\beta+c(1+z)^{3}-\frac{\gamma}{2}\bigg],
\end{equation}
The Hubble parameter expression in equation (\ref{21}) can be rewritten in a more compact and redshift-dependent form as:
\begin{equation}\label{22}
H(z)=\sqrt{H_{0}^{2}(1+z)^{3}+\frac{1}{3}\bigg(\frac{3(\alpha+\beta)}{2}(1+z)\left[(1+z)^{2}-1\right]-\beta\left[(1+z)^{3}-1\right]+\frac{\gamma}{2}\left[(1+z)^{3}-1\right]}\bigg),
\end{equation}
Here, $H_0$ represents the Hubble parameter at the current epoch. This formulation is derived by aligning equation (\ref{21}) with the standard cosmological expression and determining the integration constant $c$ by evaluating the equation at redshift $z = 0$. Enforcing the condition $H(z=0) = H_0$ allows us to explicitly solve for $c$, yielding the result as: $c=3H_{0}^{2}+\frac{3}{2}(\alpha+\beta)-\beta+\frac{\gamma}{2}$.
\section{Observational constraints and parameter estimation} \label{sec4}
\hspace{0.5cm} To evaluate the reliability and observational compatibility of our modified expansion model arising from $f(Q,\mathcal{L}_m)$ gravity, we compare its predictions against current cosmological data. The aim is to tightly constrain the model parameters—namely, the Hubble constant $H_0$, along with the phenomenological coefficients $\alpha$, $\beta$ and $\gamma$—that influence the form of the modified Hubble function $H(z)$.

For this purpose, we employ a Bayesian inference approach using the Markov Chain Monte Carlo (MCMC) method, which allows for efficient exploration of the parameter space and yields well-defined statistical estimates of the credible intervals. The MCMC analysis is carried out using the Python-based emcee package \cite{Mackey13}, which utilizes an ensemble sampler with affine-invariant transformations, enhancing the performance and convergence behavior in models with multiple free parameters. We formulate the likelihood function under the assumption that observational errors follow a Gaussian distribution. This results in a probability distribution expressed in the form:
\begin{equation}\label{23}
\mathcal{L} \propto \exp\left(-\frac{\chi^2}{2}\right),
\end{equation}
Here, $\chi^2$ represents the total chi-square statistic, which measures the degree of discrepancy between the theoretical predictions of the model and the observed data. We examine the parameter space using flat, uniform priors defined over the ranges: $50 < H_0 < 90$, $-1 < \alpha < 1$, $-1 < \beta < 1$ and  $-1 < \gamma < 1$. The MCMC simulation is carried out with $100$ walkers, each performing $2000$ steps. To ensure reliable sampling and convergence, an initial burn-in phase—determined by the autocorrelation time—is discarded. From the well-converged MCMC chains, we extract the posterior distributions of the parameters, which are visualized through marginalized one-dimensional histograms and two-dimensional contour plots. These plots display confidence intervals at the $68\%$, $95\%$ and $99.7\%$ levels, providing a clear view of the parameter constraints and correlations.

In the subsections that follow, we describe the construction of the chi-square functions for each dataset used in our analysis: (i) cosmic chronometer Hubble measurements, (ii) BAO and DESI DR2 data and (iii) Pantheon+ Type Ia supernovae sample.
\subsection{Cosmic chronometers: hubble parameter measurements} \label{sec4.1}
\hspace{0.5cm} To place observational constraints on our theoretical model, we make use of $46$ data points of the Hubble parameter $H(z)$, obtained through the cosmic chronometer (CC) method. This technique offers a model-independent approach to trace the Universe's expansion history by estimating the differential ages of passively evolving early-type galaxies across various redshifts \cite{Amits24}. By measuring how redshift changes with cosmic time, the Hubble parameter is directly determined via the relation: $H(z)=-(1+z)^{-1}\frac{dz}{dt}$. This relation arises directly from the standard FLRW cosmological framework and does not depend on any specific assumptions regarding the gravitational theory or the nature of dark energy.

The cosmic chronometer dataset provides strong and independent constraints on $H(z)$ across a broad range of redshifts, drawing from a collection of high-precision spectroscopic observations. Since this method does not rely on integrated distance indicators, it serves as a powerful and reliable tool for testing alternative gravity theories, such as the modified framework considered in this work. To assess the agreement between our theoretical model and these observations, we define the chi-square statistic for the Hubble dataset as:
\begin{equation}\label{24}
\chi^{2}_{H(z)}=\Delta H^{T}C^{-1}_{H(z)}\Delta H,
\end{equation}
Here, $\Delta H = H_{model}(z) - H_{obs}(z)$ denotes the residual vector, capturing the difference between the theoretical predictions and the observed values of the Hubble parameter at various redshifts. The matrix $C_{H(z)}$ represents the covariance matrix, which accounts for measurement uncertainties as well as any possible correlations between the observational data points.
\subsection{Baryon acoustic oscillation measurements} \label{sec4.2}
\hspace{0.5cm} Baryon Acoustic Oscillations (BAO) represent a key observational feature in cosmology, providing a fixed length scale rooted in the early Universe. This scale, known as the comoving sound horizon $r_d= \int_{z_d}^\infty \frac{c_s(z)}{H(z)} dz$, originates from acoustic waves that traveled through the tightly coupled photon–baryon fluid before the Universe cooled enough for recombination to occur. As these oscillations left an imprint on the distribution of matter, BAO measurements serve as a cosmic yardstick to trace the expansion history across time. To place constraints on our model, we employ a set of 15 up-to-date BAO measurements, including the latest high-precision data from DESI DR2 \cite{DESI24,AG25,AGA25,Karim2025}. These measurements explore the large-scale structure of the Universe through various tracers like galaxies and quasars. The observational results are presented in terms of fundamental cosmological distance indicators, such as:
\begin{equation}\label{25}
D_H(z) = \frac{c}{H(z)}, \hspace{0.4cm} D_M(z) = \int_0^z \frac{c}{H(z')} dz', \hspace{0.4cm} D_V(z) = \left[ z D_M^2(z) D_H(z) \right]^{1/3}.
\end{equation}
We compare theoretical predictions with the observed ratios $\frac{D_X}{r_d}$ for $(X=H,M,V)$ and compute the chi-square statistic:
\begin{equation}\label{26}
\chi^2_{BAO} = \Delta D^\mathrm{T} C^{-1} \Delta D,
\end{equation}
Here, $\Delta D = D_{\text{obs}} - D_{\text{th}}$ captures the deviation between observation and theory and $C^{-1}$ is the inverse covariance matrix. These data tightly constrain the Universe's expansion and geometry.
\subsection{Pantheon+ supernova dataset} \label{sec4.3}
\hspace{0.5cm} Type Ia supernovae (SNe Ia) serve as highly effective cosmic distance markers because of their consistent intrinsic brightness, making them excellent tools for tracking the Universe’s expansion history. The Pantheon+ sample, which includes $1701$ SNe Ia spanning the redshift range $z \in [0.001, 2.3]$, which stands as one of the most comprehensive collections for probing the dynamics of the late-time Universe \cite{Kowa08,RA10,Suzu12,Beto14,Scolnic18}. The measured distance modulus is directly linked to the luminosity distance $d_L(z)$ by:
\begin{equation}\label{27}
\mu(z) = 5 \log_{10} \left( \frac{d_L(z)}{1\, Mpc} \right) + 25,
\end{equation}
with $d_L(z)$ computed in a flat FLRW Universe as:
\begin{equation}\label{28}
d_L(z) = c (1 + z) \int_0^z \frac{dz'}{H(z'; H_{0},\alpha,\beta,\gamma)},
\end{equation}
To test the model, we calculate the residuals between theoretical and observed distance moduli, defining the chi-square statistic as:
\begin{equation} \label{29}
\chi^2_{\mathrm{Pantheon+}} = \Delta \mu^\mathrm{T} C^{-1}_{\mathrm{Pantheon+}}\Delta\mu,
\end{equation}
where $C_{Pantheon+}$ is the full covariance matrix and $\Delta\mu=m_{B}-M-\mu(z)$. Additionally, for Cepheid-calibrated SNe, the observed distance modulus is replaced with independently determined values to break the degeneracy between $M$ and $H_{0}$, enhancing the robustness of the constraints derived from this dataset.
\subsection{Joint likelihood analysis and parameter estimation} \label{sec4.4}
\begin{figure}[hbt!]
  \centering
  \includegraphics[scale=0.4]{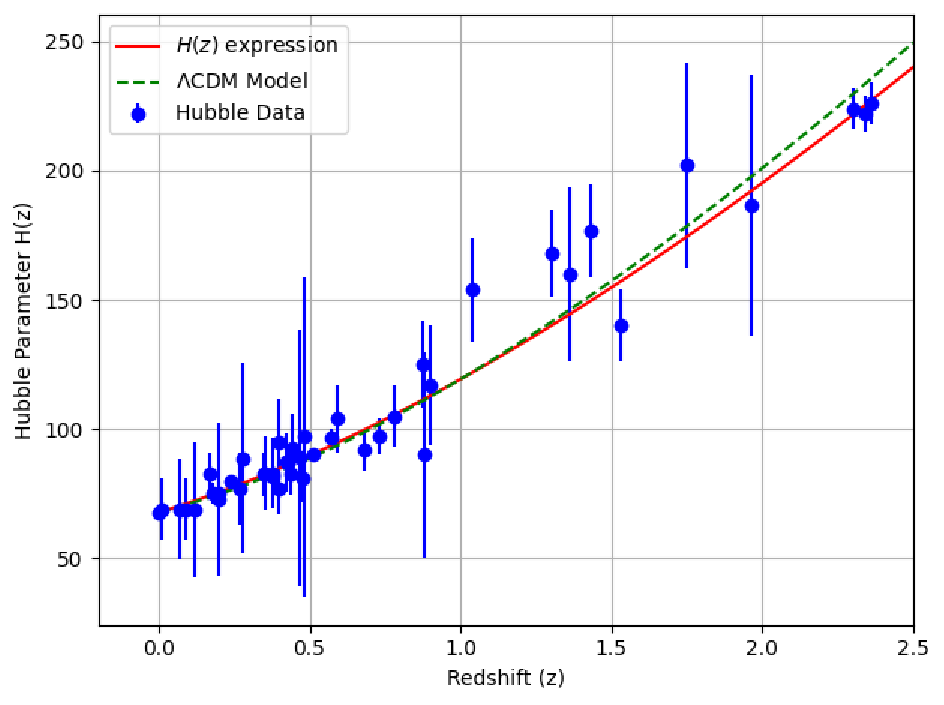}~~~~
  \includegraphics[scale=0.4]{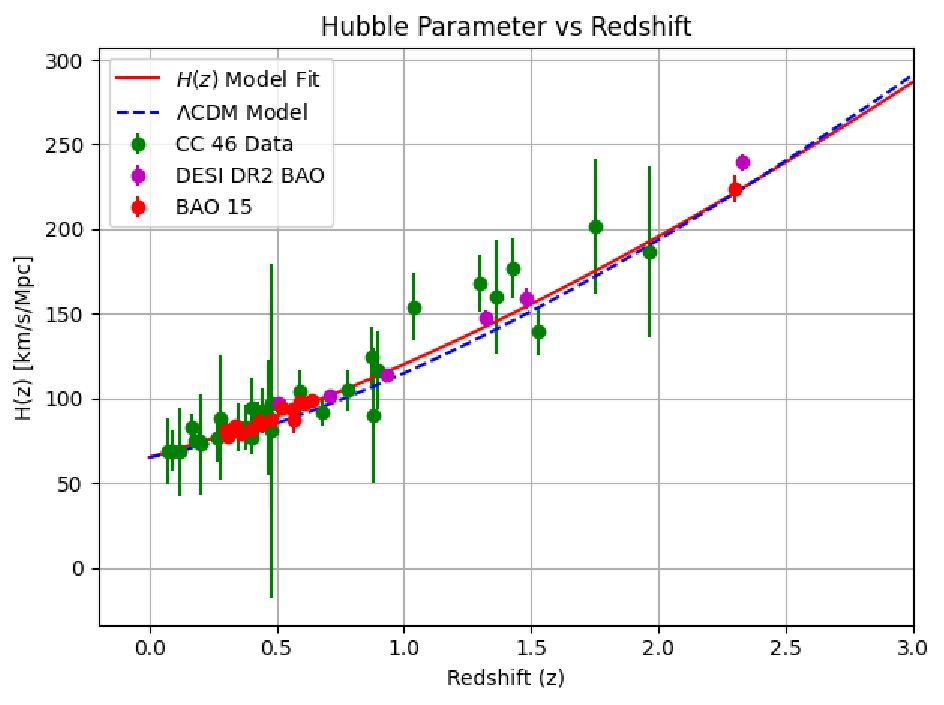}~~\\
  \includegraphics[scale=0.38]{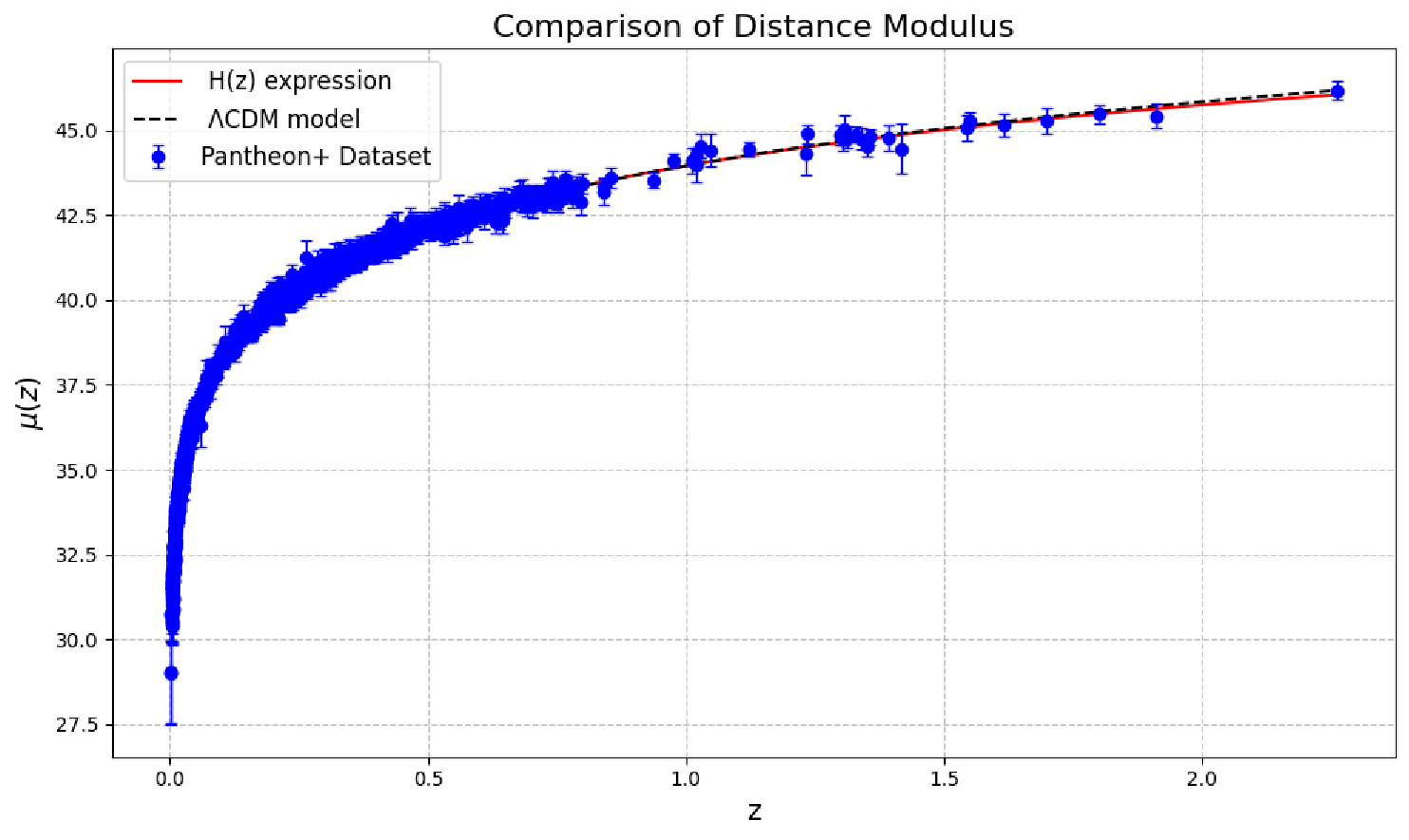}
  \caption{Error bar comparison showing how parameter estimates shift across different dataset combinations.}\label{fig:f1}
\end{figure}

\begin{figure}[hbt!]
  \centering
  \includegraphics[scale=0.42]{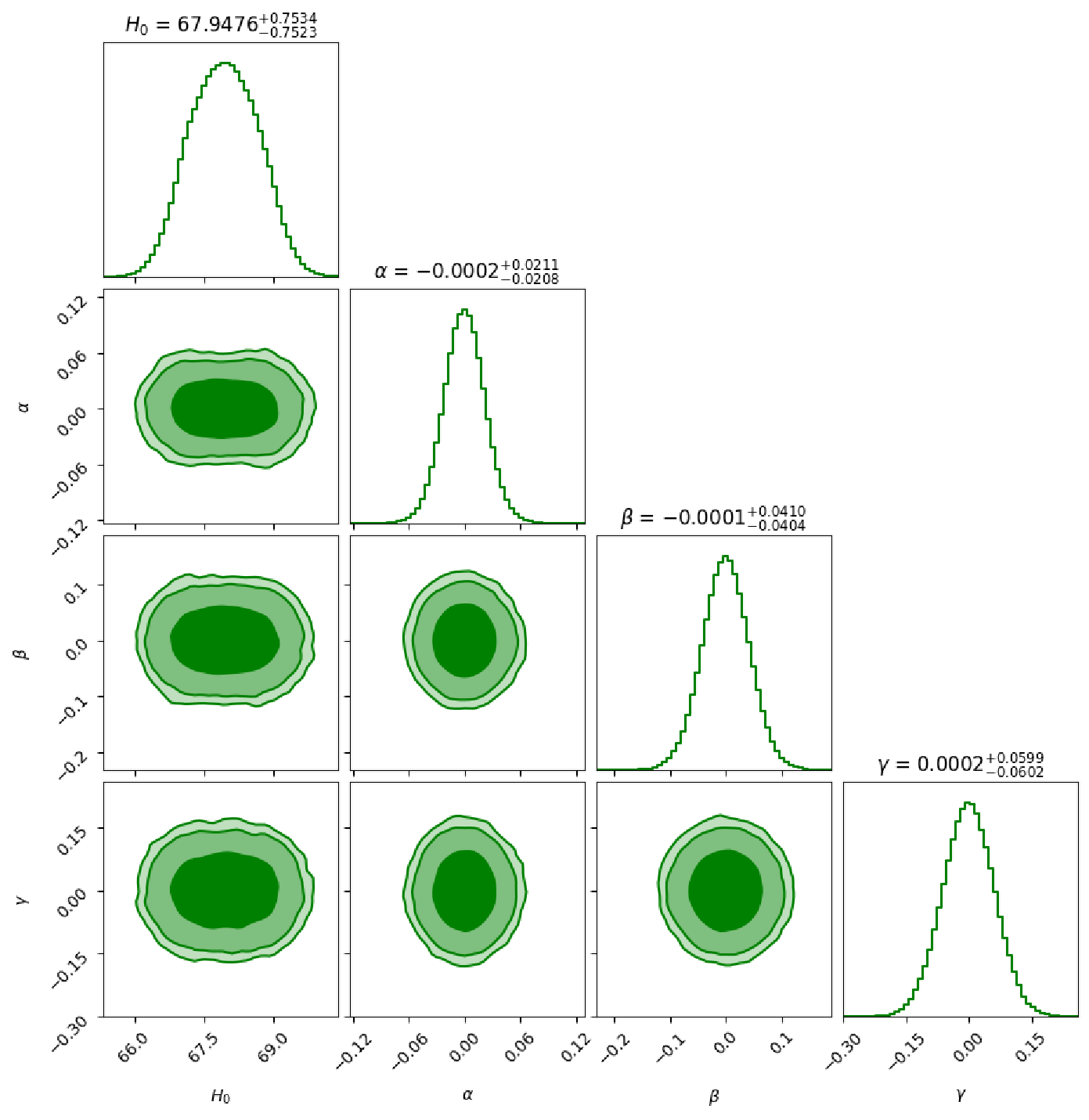}
  \caption{Joint constraint contours for $(H_{0},\alpha,\beta,\gamma)$ illustrating $1\sigma$, $2\sigma$ and $3\sigma$ confidence levels based on the combined observational datasets.}\label{fig:f2}
\end{figure}

\hspace{0.5cm} Our statistical analysis relies on a multi-faceted dataset, encompassing Hubble parameter measurements, BAO data points and a large collection of supernovae observations. By combining these datasets, we aim to precisely constrain our cosmological model's parameters and gain deeper insights into cosmic evolution. The total chi-square function, used as the basis for the likelihood analysis, is defined as:
\begin{equation}\label{30}
\chi^{2}_{total}=\chi^{2}_{H(z)}+\chi^{2}_{BAO}+\chi^{2}_{SNe Ia},
\end{equation}
and the corresponding likelihood is evaluated through a Bayesian framework:
\begin{equation} \label{31}
\mathcal{L} \propto \exp\left(-\frac{\chi^2_{\text{total}}}{2}\right).
\end{equation}
The contour plots illustrate confidence regions at the $68.27\%$ $(1\sigma)$, $95.45\%$ $(2\sigma)$ and $99.73\%$ $(3\sigma)$ levels, visually highlighting the relationships and correlations between parameters. By combining multiple datasets, this analysis enhances the precision of parameter constraints and helps to minimize degeneracies that commonly affect analyses based on a single observational source. Figures \ref{fig:f1} and \ref{fig:f2} display the resulting one-dimensional marginalized distributions and two-dimensional confidence contours.
\subsection{Best-fit values and implications of the model}\label{sec4.5}
\hspace{0.5cm} Using MCMC sampling, we derive the best-fit values for the model parameters based on the combined analysis of Hubble parameter measurements, BAO data and the Pantheon+ Type Ia supernova compilation as: $H_{0}=67.9476^{+0.7534}_{-0.7523}$ (km/s/Mpc), $\alpha=-0.0002^{+0.0211}_{-0.0208}$, $\beta=-0.0001^{+0.0410}_{-0.0404}$ and $\gamma=0.0002^{+0.0599}_{-0.0602}$. The findings show that our model yields a value of the Hubble constant $H_0$ that falls comfortably within the bounds established by both the Planck 2018 CMB data ($H_0 = 67.4 \pm 0.5$ km/s/Mpc) \cite{Planck2018} and the SH0ES local distance ladder measurements ($H_0 = 73.04 \pm 1.04$ km/s/Mpc) \cite{Riess2022}. This agreement underscores the model’s ability to accommodate both early- and late-universe observations, which offers a promising avenue for alleviating the current Hubble tension. The parameters $\alpha$ and $\beta$, which govern the redshift-dependent pressure in the chosen form $p(z)=\alpha+\frac{\beta z}{1+z}$, are both found to be consistent with zero within $1\sigma$ confidence levels. This indicates that the effective pressure exhibits minimal evolution with redshift, approaching a nearly constant value at low redshifts—behavior that resembles that of a cosmological constant. Nevertheless, the relatively broad credible interval for $\beta$ leaves room for mild deviations from the standard $\Lambda$CDM model, suggesting the possibility of dynamical dark energy behavior, particularly at intermediate redshift ranges. The inclusion of the new parameter $\gamma$, which encapsulates matter-geometry coupling in the $f(Q,\mathcal{L}_m)$ framework, is also consistent with zero within $1\sigma$, yet the credible interval allows for subtle deviations that could signal new gravitational effects.

In summary, the agreement between the model predictions and various observational datasets, along with the derived parameter values, suggests that the proposed $f(Q,\mathcal{L}_m)$ framework—with minimal coupling and a redshift-dependent pressure—stands as a compelling alternative to the conventional $\Lambda$CDM cosmology. It successfully captures a broad spectrum of cosmological dynamics while remaining within the limits set by current observations, and also provides flexibility to investigate potential departures from the standard model in the late Universe.
\section{Cosmological parameter dynamics}\label{sec5}
\subsection{Deceleration parameter}\label{sec5.1}
\hspace{0.5cm} The deceleration parameter $q(z)$ is a key cosmological indicator that helps describe the nature of the Universe’s expansion dynamics. The deceleration parameter for our model is derived from the Hubble parameter $H(z)$ provided in
\begin{equation}\label{32}
q=-1+\frac{3H_{0}^{2}(1+z)^{3}+\frac{\alpha+\beta}{2}(1+z)\left[3(1+z)^{2}-1\right]-\beta(1+z)^{3}+\frac{\gamma}{2}(1+z)^{3}}
{2\bigg[H_{0}^{2}(1+z)^{3}+\frac{1}{3}\left(\frac{3(\alpha+\beta)}{2}(1+z)\left[(1+z)^{2}-1\right]-\beta\left[(1+z)^{3}-1\right]+\frac{\gamma}{2}\left[(1+z)^{3}-1\right]\right)\bigg]}.
\end{equation}
\begin{figure}[hbt!]
  \centering
  \includegraphics[scale=0.44]{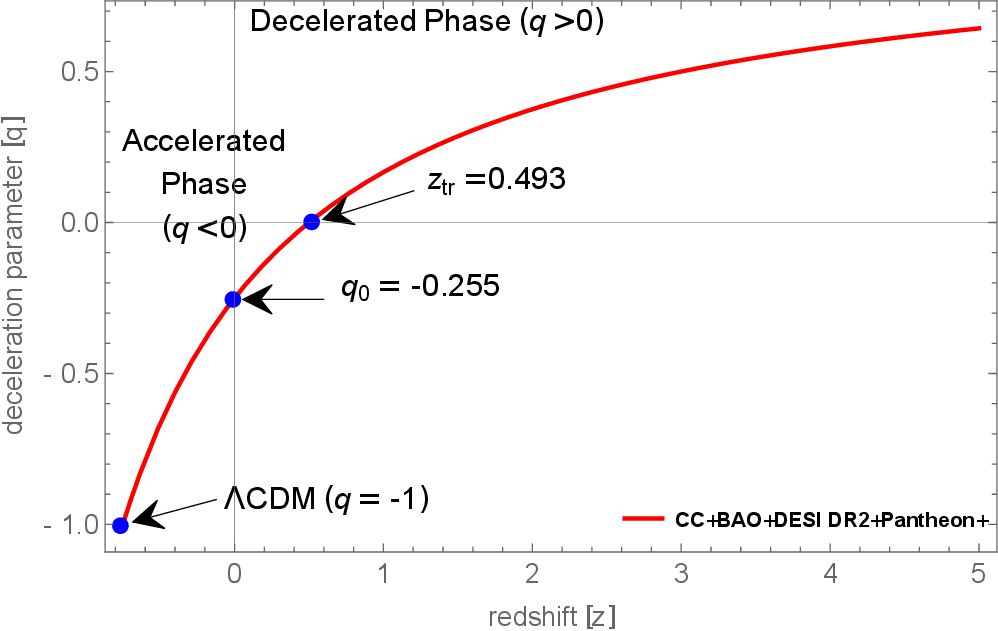}
  \caption{Redshift evolution of the deceleration parameter $q(z)$ for the proposed model.}\label{fig:f3}
\end{figure}

The redshift-dependent behavior of $q(z)$ is depicted in Figure \ref{fig:f3}. At high redshift, the curve starts with $q\approx0.645$, which reflects a phase of decelerated expansion where matter dominates the dynamics of the Universe. As cosmic evolution progresses, the deceleration parameter changes sign around the transition redshift $z_{tr}\approx0.493$, signaling the beginning of the accelerated expansion phase of the Universe. In the far future, the deceleration parameter asymptotically approaches $q \to -1$, which indicates a transition toward a de Sitter–type accelerated expansion phase. At the current epoch ($z = 0$), the model yields a deceleration parameter of approximately $q_0 \approx -0.255$, which is in good agreement with present-day observational estimates.
\subsection{Evolution of energy density and pressure}\label{sec5.2}
\hspace{0.5cm} This section focuses on analyzing how the energy density $\rho(z)$ and pressure $p(z)$ evolve with redshift, which offers insight into the cosmic fluid's dynamical properties over time. From equations (\ref{15}) and (\ref{16}), we derive the expressions of energy density and pressure as follows:
\begin{equation}\label{33}
\rho(z)=3H^{2}+\frac{\gamma}{2}=3H_{0}^{2}(1+z)^{3}+\frac{3(\alpha+\beta)}{2}(1+z)\left[(1+z)^{2}-1\right]-\beta\left[(1+z)^{3}-1\right]+\frac{\gamma}{2}(1+z)^{3},
\end{equation}
\begin{equation}\label{34}
p(z)=-2\dot{H}-3H^{2}-\frac{\gamma}{2}=(1+z)\left[\alpha+\frac{\beta z}{1 + z}\right].
\end{equation}
\begin{figure}[hbt!]
  \centering
  \includegraphics[scale=0.42]{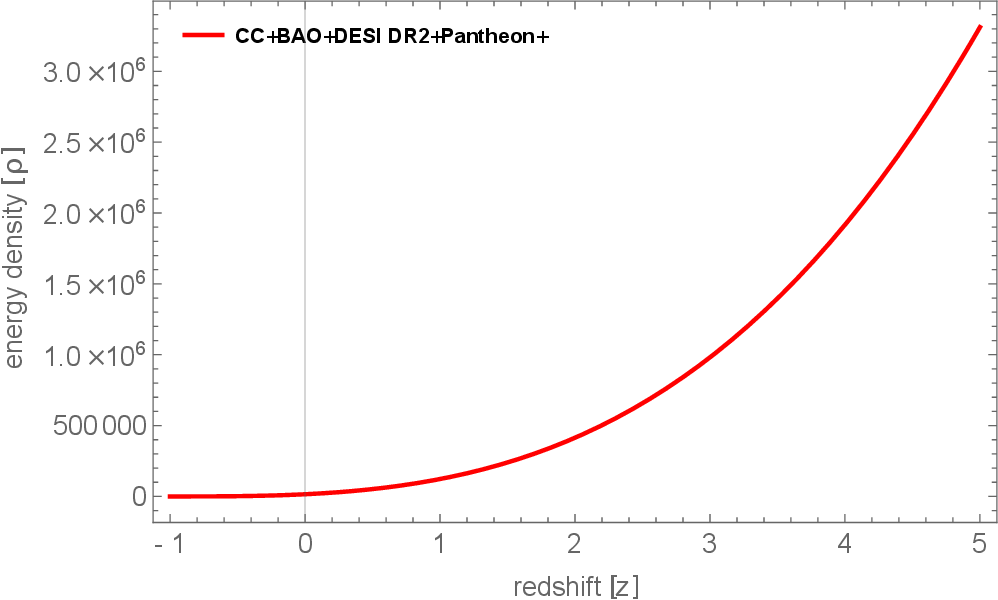}~~~
  \includegraphics[scale=0.42]{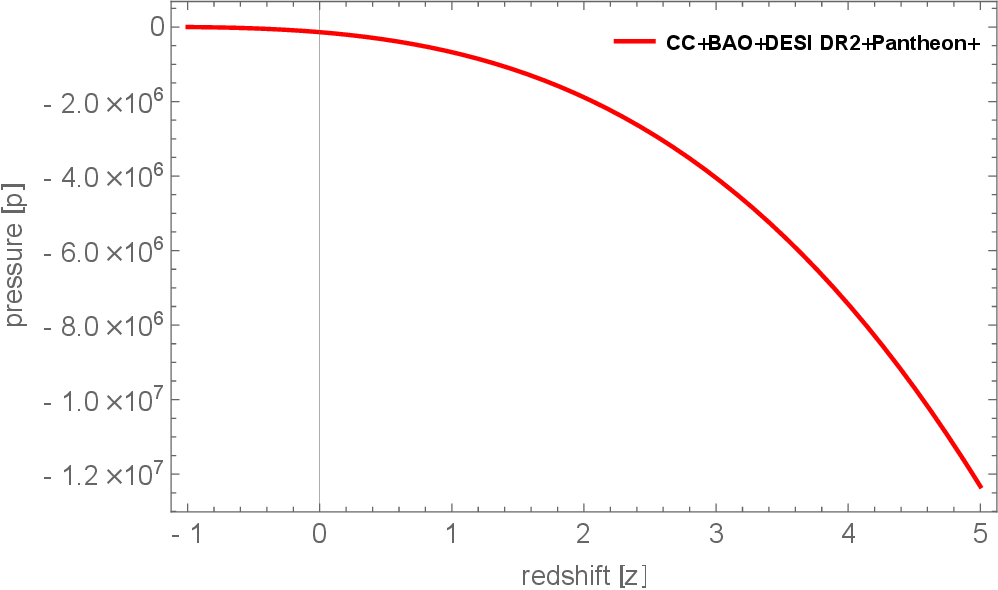}
  \caption{Redshift evolution of energy density and pressure for our model with $\gamma=0.5$.}\label{fig:f4}
\end{figure}

Using the derived expressions from our model, we find that the energy density starts from a positive value at high redshift and smoothly declines toward zero, which reflects the dilution due to cosmic expansion. Conversely, the pressure begins with a negative value—supporting cosmic acceleration—and gradually approaches zero in the far future, which signals a transition toward a pressureless state. Notably, this behavior aligns with current cosmological observations: the positive energy density at early times supports structure formation, while the negative pressure at late times is consistent with the observed accelerated expansion of the Universe. 
\subsection{Equation of state parameter}\label{sec5.3}
\hspace{0.5cm} Next, we investigate the redshift evolution of the equation of state (EoS) parameter $\omega(z)=\frac{p(z)}{\rho(z)}$, as predicted by our theoretical model. The expression of $\omega(z)$ is derived by using equations (\ref{33}) and (\ref{34}) as:
\begin{equation}\label{35}
\omega(z)=\frac{p}{\rho}=\frac{(1+z)\left[\alpha+\frac{\beta z}{1 + z}\right]}{3H_{0}^{2}(1+z)^{3}+\frac{3(\alpha+\beta)}{2}(1+z)\left[(1+z)^{2}-1\right]-\beta\left[(1+z)^{3}-1\right]+\frac{\gamma}{2}(1+z)^{3}}.
\end{equation}
\begin{figure}[h!]
  \centering
  \includegraphics[scale=0.45]{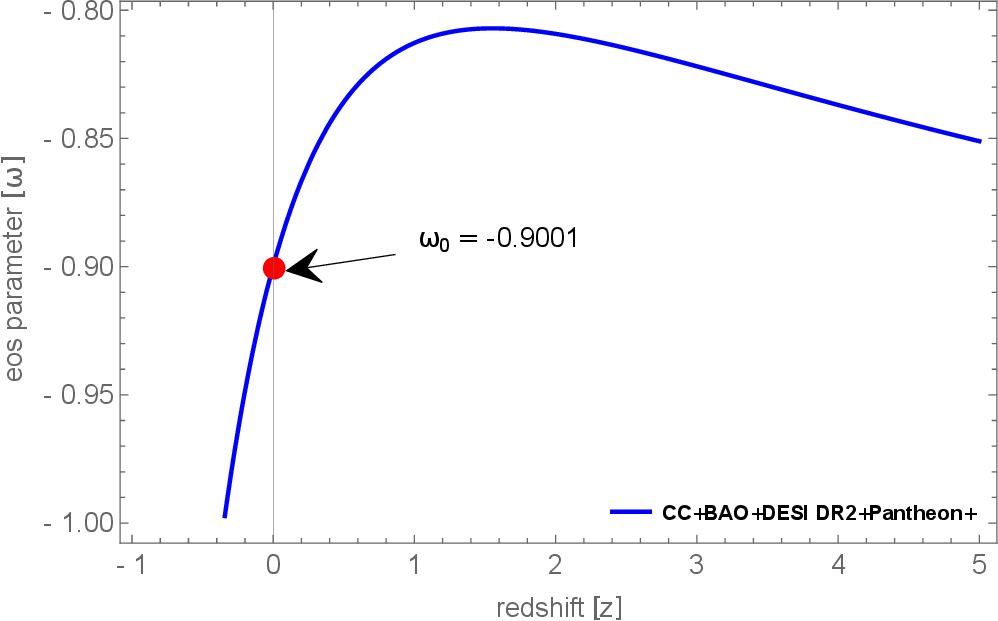}
  \caption{Evolution of $\omega(z)$ for our model with $\gamma=0.5$.}\label{fig:f5}
\end{figure}

Figure \ref{fig:f5} illustrates the behavior of $\omega$ across redshift. The plot reveals that the EoS parameter starts around $\omega \approx -0.85$ at high redshift and gradually trends toward $\omega \to -1$ in the late-time Universe. the current value is $\omega_{0}=-0.9001$, which lies well within the dark energy regime. This evolution indicates a quintessence-like phase in the early Universe, gradually transitioning into a cosmological constant–dominated era. At all redshifts, $\omega(z)<0$, which confirms a persistent negative pressure and thus supporting the observed accelerated expansion throughout cosmic history. The smooth interpolation from $\omega\approx-0.85$ to $\omega\approx-1$ also ensures that the model remains observationally viable and consistent with constraints from Planck 2018 and SNe Ia datasets.
\section{Energy conditions analysis} \label{sec6}
\begin{figure}[h!]
  \centering
  \includegraphics[scale=0.42]{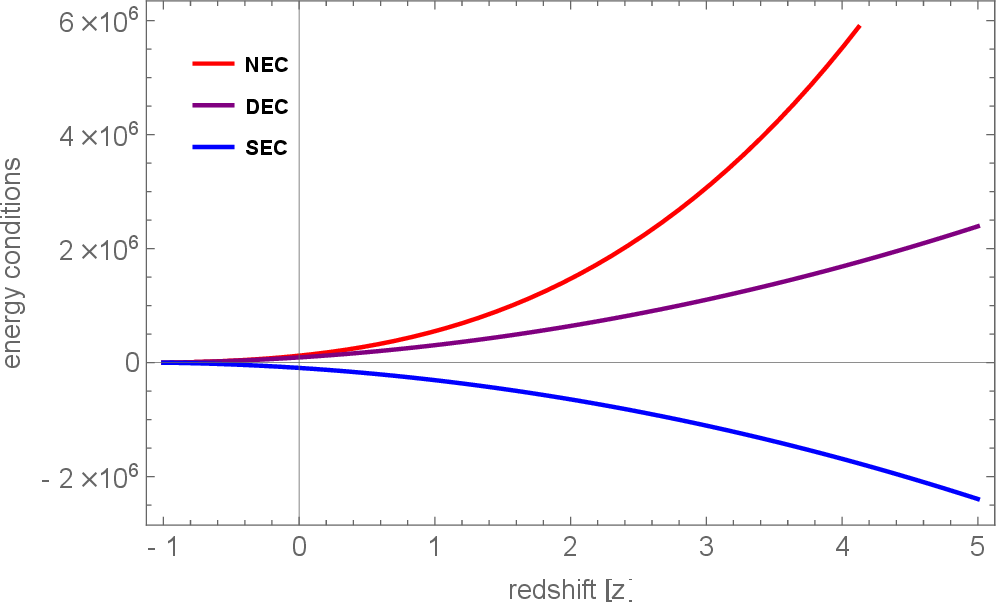}
  \caption{Evolution of energy condition expressions with redshift.}\label{fig:f6}
\end{figure}

\hspace{0.5cm} This section focuses on evaluating the standard energy conditions to test the physical consistency of our cosmological model. These conditions—namely the Null Energy Condition (NEC), Dominant Energy Condition (DEC) and Strong Energy Condition (SEC)—are formulated using the energy density $\rho$ and pressure $p$ as described by: NEC: $\rho+p\geq0$, DEC: $\rho-p\geq0$ and SEC: $\rho+3p\geq0$. Employing the obtained expressions for $\rho(z)$ and $p(z)$ from equations (\ref{33}) and (\ref{34}), we evaluate the specific combinations required to test each of the aforementioned energy conditions. Figure \ref{fig:f6} presents the redshift-dependent evolution of these quantities, which illustrates how each energy condition varies throughout cosmic history.

\begin{eqnarray}\label{36}
\rho+p&=&3H_{0}^{2}(1+z)^{3}+\frac{3(\alpha+\beta)}{2}(1+z)\left[(1+z)^{2}-1\right]-\beta\left[(1+z)^{3}-1\right]+\frac{\gamma}{2}(1+z)^{3}\\\nonumber
&&+(1+z)\left[\alpha+\frac{\beta z}{1 + z}\right],
\end{eqnarray}
\begin{eqnarray}\label{37}
\rho-p&=&3H_{0}^{2}(1+z)^{3}+\frac{3(\alpha+\beta)}{2}(1+z)\left[(1+z)^{2}-1\right]-\beta\left[(1+z)^{3}-1\right]+\frac{\gamma}{2}(1+z)^{3}\\\nonumber
&&-(1+z)\left[\alpha+\frac{\beta z}{1 + z}\right],
\end{eqnarray}
\begin{eqnarray}\label{38}
\rho+3p&=&3H_{0}^{2}(1+z)^{3}+\frac{3(\alpha+\beta)}{2}(1+z)\left[(1+z)^{2}-1\right]-\beta\left[(1+z)^{3}-1\right]+\frac{\gamma}{2}(1+z)^{3}\\\nonumber
&&+3(1+z)\left[\alpha+\frac{\beta z}{1 + z}\right],
\end{eqnarray}

The plotted results reveal that the NEC and DEC are satisfied across the entire redshift range, which suggests that the energy-momentum content of the model remains physically reasonable, which avoids unphysical behaviors such as superluminal propagation or negative-energy instabilities. In contrast, the SEC is violated which is a characteristic outcome commonly associated with accelerated expansion in the late Universe. This violation aligns with expectations for models aiming to describe the current phase of cosmic acceleration.
\section{Slow-Roll dynamics in the context of $f(Q,\mathcal{L}_m)$ gravity}\label{sec7}
\hspace{0.5cm} In inflationary cosmology, the slow-roll approximation plays a fundamental role in describing how the scalar field evolves during the early phase of rapid expansion. Within our framework, analyzing the slow-roll parameters offers valuable understanding of how inflationary dynamics emerge from the modified gravity model and how effectively they correspond to a scalar field-driven scenario. The first and second slow-roll parameters related to the Hubble expansion are introduced as follows \cite{Shir22,Liddle94}:
\begin{equation}\label{39}
\epsilon_{1}=-\frac{\dot{H}}{H^{2}}, \hspace{0.6cm} \epsilon_{2}=\frac{\ddot{H}}{H\dot{H}}-2\frac{\dot{H}}{H^{2}}.
\end{equation}
which are directly computed using our model's Hubble function derived in equation (\ref{22}) in the following way:
\begin{equation}\label{40}
\epsilon_{1}=\frac{3H_{0}^{2}(1+z)^{3}+\frac{\alpha+\beta}{2}(1+z)\left[3(1+z)^{2}-1\right]-\beta(1+z)^{3}+\frac{\gamma}{2}(1+z)^{3}}
{2\bigg[H_{0}^{2}(1+z)^{3}+\frac{1}{3}\left(\frac{3(\alpha+\beta)}{2}(1+z)\left[(1+z)^{2}-1\right]-\beta\left[(1+z)^{3}-1\right]+\frac{\gamma}{2}\left[(1+z)^{3}-1\right]\right)\bigg]},
\end{equation}
\begin{eqnarray}\label{41}
\epsilon_{2}&=&-\frac{9H_{0}^{2}(1+z)^{3}+\frac{\alpha+\beta}{2}(1+z)\left[9(1+z)^{2}-1\right]-3\beta(1+z)^{3}+\frac{3\gamma}{2}(1+z)^{3}}
{3H_{0}^{2}(1+z)^{3}+\frac{\alpha+\beta}{2}(1+z)\left[3(1+z)^{2}-1\right]-\beta(1+z)^{3}+\frac{\gamma}{2}(1+z)^{3}}\\\nonumber
&&+\frac{3H_{0}^{2}(1+z)^{3}+\frac{\alpha+\beta}{2}(1+z)\left[3(1+z)^{2}-1\right]-\beta(1+z)^{3}+\frac{\gamma}{2}(1+z)^{3}}
{H_{0}^{2}(1+z)^{3}+\frac{1}{3}\left(\frac{3(\alpha+\beta)}{2}(1+z)\left[(1+z)^{2}-1\right]-\beta\left[(1+z)^{3}-1\right]+\frac{\gamma}{2}\left[(1+z)^{3}-1\right]\right)}.
\end{eqnarray}
\begin{figure}[h!]
  \centering
  \includegraphics[scale=0.42]{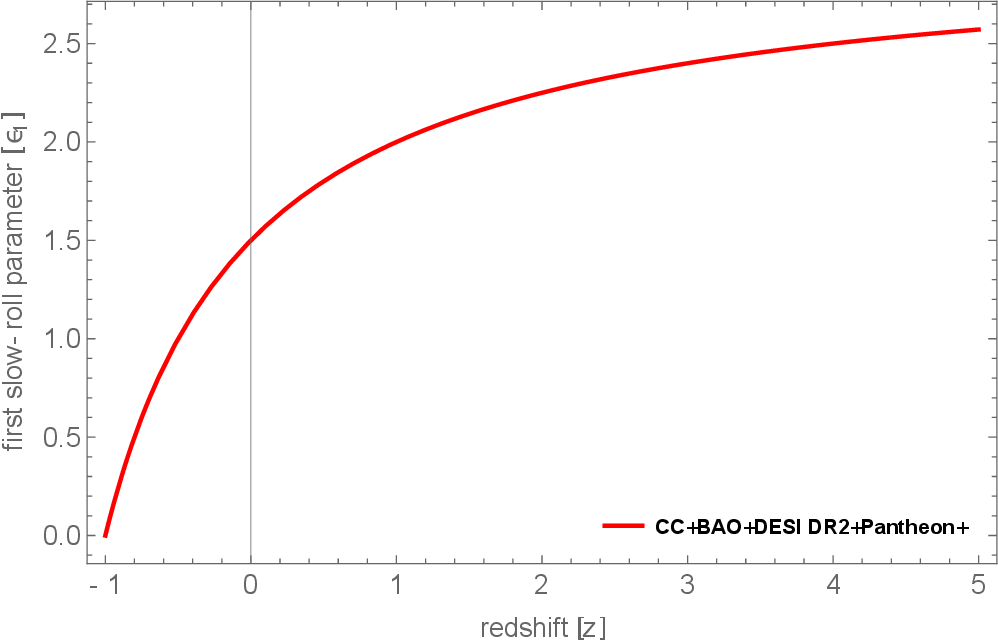}~~~
  \includegraphics[scale=0.42]{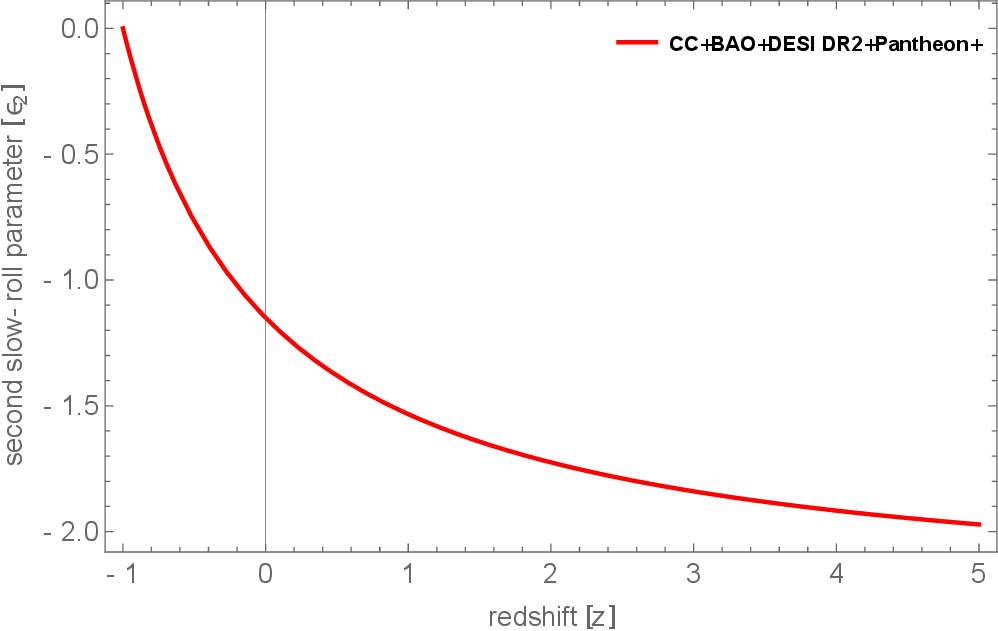}
  \caption{Evolution of the slow-roll parameters $\epsilon_{1}(z)$ and $\epsilon_{2}(z)$ for the proposed $f(Q,\mathcal{L}_m)$ model.}\label{fig:f7}
\end{figure}

Figure \ref{fig:f7} presents the redshift evolution of the slow-roll parameters $\epsilon_1$ and $\epsilon_2$, providing insight into inflationary behavior within our $f(Q,\mathcal{L}_m)$ framework. The first parameter $\epsilon_1$ initially takes a large value around $2.5$, gradually declines over time, and reaches the critical threshold $\epsilon_1 = 1$ at approximately $z = -0.48$, signifying the termination of the inflationary phase. As the Universe evolves further, $\epsilon_1$ continues to decrease and approaches zero, which signals a transition into a phase of sustained accelerated expansion. At the same time, the second slow-roll parameter $\epsilon_2$ starts from approximately $-1.8$ and progressively rises toward zero as redshift decreases. This trend reflects a gradual and stable change in the first parameter $\epsilon_1$, maintaining the condition $|\epsilon_2| \ll 1$ throughout the slow-roll phase, which confirms the persistence of slow and steady inflationary dynamics. These findings are consistent with theoretical predictions of inflation and align with observational constraints from datasets like Planck $2018$, which favor models where $\epsilon_1 \ll 1$ and $|\epsilon_2|$ remains small throughout inflation. The evolution patterns of $\epsilon_1$ and $\epsilon_2$ in this scenario validate that the model effectively captures an inflationary phase that naturally transitions into the Universe’s current phase of accelerated expansion.
\section{Conclusion}\label{sec9}
\hspace{0.5cm} This work explored the cosmological consequences of a modified gravity theory based on the $f(Q,\mathcal{L}_m)$ formalism. Specifically, we focused on a linear functional form given by $f(Q,\mathcal{L}_m)=-Q+2\mathcal{L}_m+\gamma$, which introduces a minimal yet nontrivial coupling between non-metricity and matter, along with an effective constant term. To describe the late-time cosmic evolution, we employed a pressure parametrization dependent on redshift, given by $p(z)=\alpha+\frac{\beta z}{1+z}$. This form allowed us to derive an explicit analytical solution for the Hubble parameter $H(z)$, which captures the expansion history within the framework of the proposed modified gravity model. We carried out an extensive parameter analysis by comparing the model’s predictions with multiple up-to-date cosmological observations. This included $46$ data points from Hubble parameter measurements, $15$ BAO observations—incorporating the latest DESI DR2 results—and the Pantheon+ compilation of Type Ia supernovae. Employing the MCMC technique, we derived constraints on the model parameters, which yields the best-fit estimates: $H_{0}=67.9476^{+0.7534}_{-0.7523}$ (km/s/Mpc), $\alpha=-0.0002^{+0.0211}_{-0.0208}$, $\beta=-0.0001^{+0.0410}_{-0.0404}$ and $\gamma=0.0002^{+0.0599}_{-0.0602}$. These values are consistent with existing observational limits, which indicates strong agreement between the model and current cosmological data.

Our investigation of the deceleration parameter $q(z)$ revealed a cosmic transition point, where the Universe shifts from a decelerating regime to an accelerating phase around the redshift $z_{tr} \approx 0.493$. In the early Universe, the model predicts a deceleration parameter of approximately $q \approx 0.645$, characteristic of matter-dominated expansion. At the current epoch, it estimates $q_0 \approx -0.255$, which aligns with observational data and indicating the Universe is presently undergoing accelerated expansion.

The behavior of energy density and pressure over redshift reveals that the energy density begins with a positive value and steadily declines as the Universe expands, eventually approaching zero. Meanwhile, the pressure stays negative throughout and slowly tends toward zero in the distant future. The EoS parameter $\omega(z)$ begins near $-0.85$ and evolves towards $-1$, with the present dat value $\omega_{0}=-0.9001$, which confirms a quintessence-like dark energy behavior throughout the cosmic history. 

Our analysis of the classical energy conditions shows that both the NEC and DEC hold across the entire redshift interval. In contrast, the SEC is violated at low redshifts, a behavior commonly associated with accelerated cosmic expansion. The analysis of the slow-roll parameters, $\epsilon_1$ and $\epsilon_2$, shows that the model permits an initial inflationary phase, where $\epsilon_1$ starts around $2.5$ and drops below unity near $z=-0.48$, signaling the end of inflation and the onset of acceleration. Meanwhile, $\epsilon_2$ evolves toward zero at late times, supporting a sustained period of accelerated cosmic growth.

Overall, the model we considered is found to be consistent with observational data and displays a unified description of cosmic evolution — from early deceleration, through the transition phase, to the current acceleration. Its agreement with scalar field dynamics, energy conditions and slow-roll inflation criteria makes it a viable candidate for describing the late-time behavior of the Universe in the context of $f(Q,\mathcal{L}_m)$ gravity.

\end{document}